\documentclass[RNAAS]{aastex631}

\usepackage{verbatim, amsmath, amsfonts, amssymb,amssymb}
\usepackage{multirow}
\usepackage[utf8]{inputenc}
\usepackage{empheq}
\usepackage[skins,theorems]{tcolorbox}
\usepackage{latexsym}
\usepackage{bigints}
\usepackage{natbib}
\usepackage{color}
\usepackage{lipsum}
\usepackage{tikz}
\usepackage{fontawesome}
\usepackage{hyperref}

\newcommand{\github}[1]{%
   \href{#1}{\faGithub}%
}
\shorttitle{How have astronomers cited other fields in the last decade}
\shortauthors{Delli Veneri et al.}
\graphicspath{{./}{figures/}}
\begin{document}

\title{How have astronomers cited other fields in the last decade?}

\correspondingauthor{Rafael S. de Souza}
\email{drsouza@shao.ac.cn}
%\correspondingauthor{Michele Delli Veneri}
%\email{michele.delliveneri@unina.it}

\author[0000-0002-8178-2942]{Michele Delli Veneri}
\affiliation{University of Naples Federico II, DIETI, Naples, Italy}
\author[0000-0001-7207-4584]{Rafael S. de Souza}
\affiliation{Shanghai Astronomical Observatory,
Chinese Academy of Sciences, Shanghai 200030, China}
\author[0000-0002-2308-6623]{Alberto Krone-Martins}
\affiliation{Donald Bren School of Information and Computer Sciences, University of California, Irvine, CA 92697, USA}
\affiliation{CENTRA/SIM, Faculdade de Ci\^{e}ncias, Universidade de Lisboa, Ed. C8, Campo Grande, 1749-016, Lisboa, Portugal}
\author[0000-0002-0406-076X]{E. E. O. Ishida}
\affiliation{Université Clermont Auvergne, CNRS/IN2P3, LPC, F-63000 Clermont-Ferrand, France}
\author[0000-0002-1178-8169]{M.~L.~L.~Dantas}
\affiliation{Nicolaus Copernicus Astronomical Center, Polish Academy of Sciences, ul. Bartycka 18, 00-716, Warsaw, Poland}
\author{Noble Kennamer}
\affiliation{Donald Bren School of Information and Computer Sciences, University of California, Irvine, CA 92697, USA}
\collaboration{6}{for the COIN collaboration}

%% can be used either before or after the list of corresponding authors. The iiii

\begin{abstract}
We present a citation pattern analysis between astronomical papers and 13 other disciplines, based on the \texttt{arXiv} database over the past decade ($2010 - 2020$). We analyze 12,600 astronomical papers citing over 14,531 unique publications outside astronomy. Two striking patterns are unraveled. First, general relativity recently became the most cited field by astronomers, a trend highly correlated with the discovery of gravitational waves. Secondly, the fast growth of referenced papers in computer science and statistics, the first with a notable 15-fold increase since 2015. Such findings confirm the critical role of interdisciplinary efforts involving astronomy, statistics, and computer science in recent astronomical research.  
\end{abstract}



\section{Introduction} 
The citation patterns of scholarly works provide critical information for understanding how knowledge flows between scientific fields. These patterns naturally indicate the existence of some intellectual influence from the cited work on the paper that originates the citation \citep[e.g.][]{10.1214/ss/1177010655}, providing a way to trace how scientists communicate \citep{Martin2008, Okamura2019}. Although individual citations of single papers cannot easily quantify the influence between fields, an analysis of large publication samples can provide a quantitative signal of this influence \citep[e.g.][]{Palacios2004}. This work analyzes how astronomy and astrophysics cited other disciplines in the past decade.

\section{Data}

To analyze the citation patterns from the astronomical production, we adopt paper pre-prints available in the \texttt{arXiv} repository as a proxy for the field production. We use the \texttt{unarXive} data set~\citep{Saier2020}, a freely available bibliographic database containing over one million full-text documents, links to $2.7$ million unique publications through $15.9$ million references, and $29.2$ million citations published between $2007$ and $2020$. We built the dataset through an automatic pipeline that uses \texttt{arXiv} and \texttt{Microsoft Academic Graph} \citep{sinha2015overview} to recover documents and map the citations. We restricted this analysis to the freely accessible subset of the data using the \texttt{arXiv} API, thus making this study reproducible. Our final sample contains $3,800,968$ references connecting $611,510$ unique scientific articles, published between $2010$ and $2020$ and spawning all $8$ primary categories of the \texttt{arXiv} classification taxonomy, as well as several subcategories. For our analysis, the $152$ available categories were combined into the following $14$ macro-categories: astronomy, biology, electrical engineering, high energy physics, nuclear physics, statistics, computer science, finance, mathematics, physics (containing all other categories of physics), condensed matter, general relativity, nonlinear sciences, and quantum physics.

\section{Analysis and Discussion}

Using the above data, we calculate citations from astronomy to other disciplines. As a first step, we count how frequently an astronomical paper cites another one outside astronomy. Then, we aggregate these counts by year of the citing papers and normalize them by the total number of papers cited in that year. This procedure quantifies a proxy measure for the relative influence of the cited fields within each year \citep[e.g.][]{Palacios2004}, and the evolution of this influence throughout the years. Finally, we examine the time evolution of the citation pattern between astronomy and other disciplines in terms of the relative proportion of the cited papers.

\begin{figure}[ht!]
\plotone{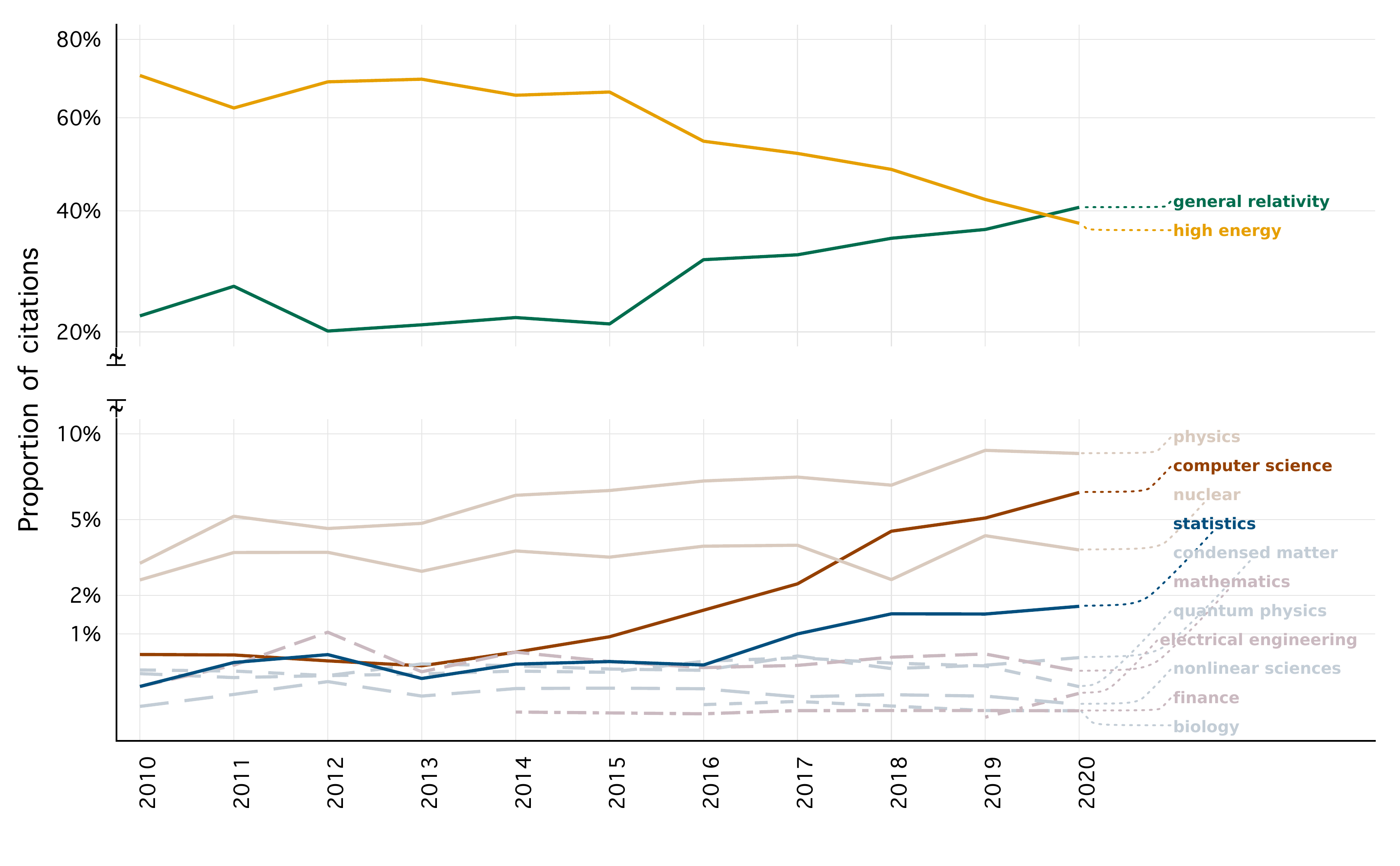}
\caption{Normalized citations from astronomy to the 13 other core disciplines in the 2010-2020 period. We highlight disciplines with noteworthy trends, including the reversal of influence between general relativity and high-energy physics and the fast pace growth of computer science and statistics. 
\label{fig1}
}
\end{figure}

\autoref{fig1} shows noteworthy trends. The normalized citation rate from astronomy to most fields remained stable. However, since  2015 references to general relativity have experienced a 1.9-fold relative increase. This trend strongly correlates with the first announcement of the detection of gravitational waves \citep{PhysRevLett.116.061102}. Recently, papers tagged as general relativity surpassed high-energy physics as the dominant field in the outwards astronomical citation flow. The citation patterns to computer sciences and statistics indicate another interesting behavioral change. Starting around 2013, the proportion of astronomical papers citing computer sciences steadily increased, climbing from 0.43\% to about 6.4\%, an impressive 15-fold growth. Surprisingly, the proportion of citations to the computer sciences field is growing at a pace that might surpass general physics (including traditional fields such as optics and instrumentation), trailing only general relativity and high-energy physics. Finally, citations from astronomy to statistics have also been increasing, starting around 2016, and albeit at a slower pace than computer sciences, it might surpass nuclear physics, which has held a steady citation rate throughout the past decade. 

Astronomy is now firmly in a long-predicted era \citep[e.g.][]{2001ASPC..225.....B,2001Sci...293.2037S} where acquiring and timely analyzing enormous amounts of data requires a new statistical methodology and computational algorithms. Other disciplines face similar challenges, and a plethora of methodology, software, and computational-thinking education approaches can be shared, thus making cross-disciplinary awareness an almost mandatory ingredient for innovative scientific development in the 21$^{\rm st}$ century. 
This Research Note used normalized citation counts as simple heuristics to quantify the intellectual influence between astronomy and other core disciplines in the hope that these results can be valuable for future decision-making resource allocation, especially those focusing on interdisciplinary projects.

\begin{acknowledgments}
RSS thanks the National Natural Science Foundation of China, grant E045191001.  AKM thanks the Portuguese Funda\c c\~ao para a Ci\^encia e a Tecnologia grants UID/FIS/00099/2019 and PTDC/FIS-AST/31546/2017. MLLD acknowledges support from the National Science Centre, Poland, project 2019/34/E/ST9/00133. The Cosmostatistics Initiative (COIN, \url{https://cosmostatistics-initiative.org/}) is an international network of researchers whose goal is to foster interdisciplinarity inspired by astronomy.
\end{acknowledgments}


\vspace{5mm}



\bibliography{ref}{}
\bibliographystyle{aasjournal}



\end{document}